\documentstyle[sprocl]{article}

\input{psfig}

\def\be{\begin{equation}}
\def\ee{\end{equation}}
\def\bea{\begin{eqnarray}}
\def\eea{\end{eqnarray}}

\begin{document}

\title{THE TRANSVERSE CROSS SECTION IN EXCLUSIVE VECTOR MESON PRODUCTION
\footnote{Presented by J.R.C. at the Fourth Workshop on Quantum Chromodynamics, 
1 - 6 June 1998, The American University of Paris}}

\author{J.R. CUDELL and I. ROYEN}

\address{
Inst. de Physique, B\^at B5, Universit\'e de Li\`ege au Sart Tilman\\
B4000 Li\`ege, Belgium\\
E-mails: JR.Cudell@ulg.ac.be, iroyen@ulg.ac.be}


\maketitle\abstracts{
We show that  the ratio of the transverse
to the longitudinal cross sections
observed at HERA cannot be reproduced by models which assume that the meson
can be described by an on-shell $q\bar q$ pair.
We explain how a simple model allowing off-shell (anti)quarks
can naturally lead to agreement with experiment.}

\section{Introduction}
Exclusive production of vector mesons at HERA has attracted a lot of attention
in recent years because of the theoretical suggestion that this
is a good way to measure the gluon distribution, as simple arguments
\cite{Ryskin}
suggest that the cross section behaves as the square of $xg(x)$.
Recently, these arguments have been formalised and extended by Collins,
Frankfurt
and Strikman \cite{Collins} in the context of a factorisation theorem for
diffractive processes. One outcome of this theorem is that the $longitudinal$
cross section (describing the transition of a longitudinal photon to a
longitudinal vector meson) is indeed proportional to the gluon distribution
squared,
and that the amplitude for the process can be described, after integration
over transverse motion,  by a convolution
in longitudinal
momentum of a meson wave function, a hard scattering amplitude, and a gluon
distribution. Furthermore, the theorem implies that this convolution is
correct at high $Q^2$ for any meson, irrespective of its mass.

Hence one has in principle a golden process to measure the gluon distribution
directly.
However, an outstanding problem has
been that the theorem holds only for longitudinal vector mesons.
The transverse part of the cross section can only be related to
quark transversity distributions, which are so far unknown.
Gauge-invariance implies
that the cross section is fully transverse in $Q^2=0$ photoproduction, but
very little else was known until now about the transverse part.

The na\"\i ve use of the gluon distribution in $\sigma_L$ as in $\sigma_T$
leads to the prediction the ratio
$R_{L/T}=\sigma_L/\sigma_T$ of longitudinal to transverse cross sections is
proportional
to $Q^2$, and hence that the transverse cross section is negligible
at high $Q^2$.
This hope has been contradicted by the observation \cite{data}  by the
ZEUS and H1 collaborations that $R_{L/T}$ was of the order of 4 at the
highest values of $Q^2$ reachable at HERA, and that furthermore it did not
seem to be linear in $Q^2$ but rather presented a high-$Q^2$ plateau,
as can be seen in Fig.~1.

\begin{figure}[t]
\vglue 2.9cm
\hglue 3.5cm (a)\hglue 5.5cm (b)\\
\vglue -3.4cm
\centerline{
\hbox{
\psfig{file=R,height=1.9in}
}\hglue 0.5cm \psfig{file=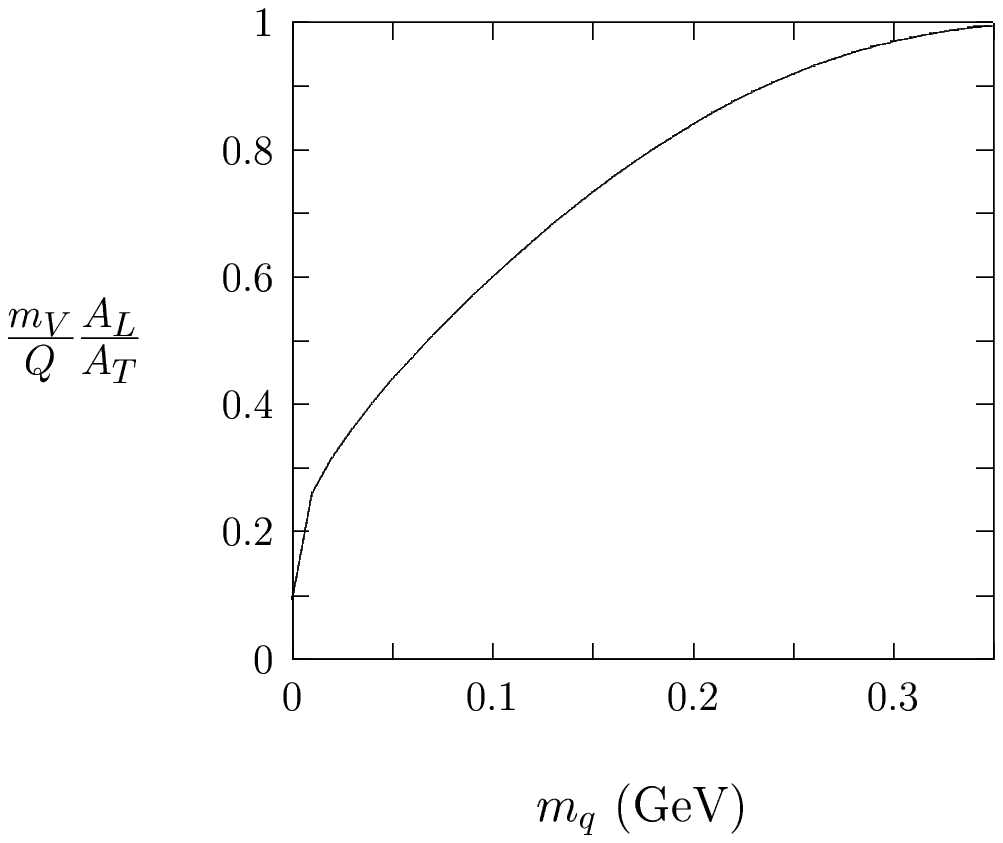,height=1.9in}}
\caption{(a) Ratio of the longitudinal and
transverse parts of the $\rho$ cross section as functions of $Q^2$;
(b) The reduction factor in the ratio $\sqrt{\sigma_L/\sigma_T}$ as a
function of the quark mass}
\end{figure}
Our goal here is to show that despite this surprise, some theoretical
handle on this process can be obtained. We shall not give all the details,
which can be found in Ref.~4, but rather show what the HERA observation mean,
and provide a simple model which reproduces the observed ratio.
\section{On-shell quarks}
One usually assumes that the exclusive process at high $s$ proceeds
from the emission of a pair of gluons from the proton, which interact with
a $q\bar q$ pair emerging from the photon, and which transfer momentum so
that this pair can be turned into a vector meson, as shown in Fig.~2.
\begin{figure}[t]
\centerline{
\psfig{file=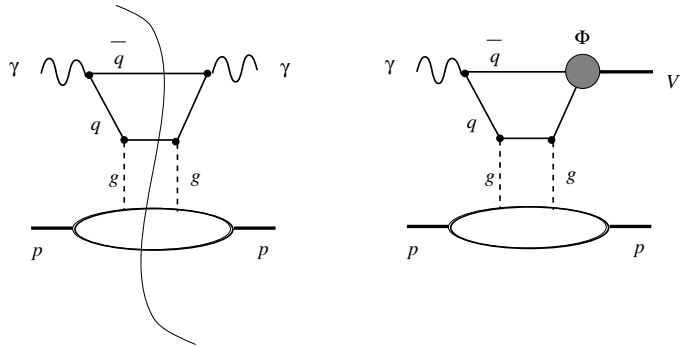,height=1.8in}
}
\caption{ The standard DIS process, compared with exclusive
vector meson production}
\end{figure}
Cutting the process, one sees
that the lower part of the diagram looks exactly as the gluon structure
function, albeit in all generality the off-diagonal one, as the process
cannot occur at $t=0$. The lower part of the diagram mainly controls
the $w^2$ and $t$ dependences of the process, whereas the upper loop
controls the helicity structure of the amplitude and its $Q^2$ dependence.
Hence the anomalous behaviour of $R_{L/T}$ has to come from the behaviour
of the upper loop, and we must look again in detail at the treatment of
the bound state of quarks.

{}From previous works \cite{Ryskin,CR1}, we already know that neglecting
Fermi motion and putting quarks on-shell lead to
$R_{L/T}=Q^2/m_V^2$, i.e. about a factor 10 above the data at the highest
$Q^2$.
One can refine the formalism further by the introduction of Fermi momentum
in the loop. A simple parton model implementation assumes that
the quark and the antiquark which make the meson are left on-shell. To examine
the consequences
of this assumption, we slightly extend our previous model \cite{CR1}
and introduce a 3-point vertex function $\Phi(l^2) \gamma_\mu$
describing the transition
$q\bar q\rightarrow V$, $l$ being the relative 4-momentum of the quarks.
As the lower part of the diagram is not of
paramount importance for the problem at hand, we treat it as a sum
of two dipole form factors, which partially kill the infrared
divergence coming from the gluon propagators of transverse impulsion $k_T$.
The amplitude is then purely imaginary,
and we use cutting rules to evaluate it.

The cutting rules do not demand that we put on-shell both $q$ and $\bar q$
from the meson.
In fact, a kinematical argument shows that such a contribution can
only be present in the case of exclusive production of an object
with positive off-shellness,
and  not in the standard DIS case of Fig.~2.a. In exclusive
meson production we can follow the usual
rules to calculate the discontinuity due to the pole in the quark propagator,
and the answer will then match the one usually calculated in a wavefunction
formalism.
Putting $q$ and $\bar q$  on-shell selects a particular value for the
quark relative momentum $l$ and for $\Phi(l^2)=\Phi_0$, and hence the exact
form of the wave-function
does not enter the ratio. At $t=0$, and for large $Q^2$, we obtain the
simple expressions:
\newcommand{\adelt}{{\bf\Delta_t}}\newcommand{\akt}{{\bf k_t}}
\newcommand{\cpt}{c_{\phi-\theta}}
\newcommand{\bl}{\beta}\newcommand{\cp}{c_\phi}
\newcommand{\alt}{{\bf
l_t}}\newcommand{\prop}{{\cal P}}
\newcommand{\mqt}{m_q^2}
 \begin{eqnarray}
dA_{L(disc)}&\approx& {-8 m_V \beta_d \Phi_0\over \akt^2 Q^3}\ {d\akt^2 \over (2
\pi)^3}\\
dA_{T(disc)}&\approx& {-4 m_V^2 \beta_d +2(\mu_q^2+2m_V^2) \log({1 + \beta_d
\over 1-\beta_d})\Phi_0\over \akt^2 Q^4}\ {d\akt^2 \over (2 \pi)^3}
\end{eqnarray}
with
$\beta_d^2={M_V^2-4m_q^2 \over M_V^2}$. These amplitudes
still need to be convoluted
in $k_T$ with the proton form factor, which will remove the remaining
infrared pole.

We see
that the ratio $ A_L/ A_T$ is still linear, but depends on the quark
mass chosen. In the limit
$\beta_d\rightarrow 0$, we recover our previous results \cite{CR1}
 for the
ratio, and in general the ratio $Q/M_V$ gets multiplied by a constant, which
depends
on the quark mass. Unfortunately, this constant is always between 0.5 and 1
for reasonable values of the (constituent) quark mass, as shown in Fig.~1.b.
The argument can be extended for nonzero $t$, with the same conclusion.
We also tried to introduce an effective (momentum-dependent) quark mass,
so that we could keep it ``on-shell'' while performing a true convolution
with the meson wavefunction. This still leads to a linear behaviour,
and does not limit the rise enough to make it compatible with the data.
\section{Off-shell quarks}
The only possibility left is to extend the model by allowing one
quark to be off-shell (the other is kept on shell by the cutting rules).
This seems to be reasonable, as it is indeed the case for the usual
DIS case of Fig.~2. There, the quark has always a negative off-shellness of
order $Q^2$. In the meson case, the situation is a little more complicated
due to the imbalanced kinematics, and one in general has a contribution
to the amplitude both from the discontinuity discussed in the previous
section, and from the principal part integral.
The behaviour of
the amplitudes is as follows, again at $t=0$ and large $Q^2$:
\begin{eqnarray}
 dA_{L(PP)}&=&{\Phi(l^2) dl^2\ d\akt^2 \over (2\pi)^3\ \akt^2\ m_V Q^3}
\times\left\{{2m_V^2 (1+\beta_+)\over \prop} \right.\nonumber\\&+&
\left.\log\left[{(4 \prop-\akt^2)\ \prop
\over (1+\beta_+)^2 Q^4}\right]
-4{\prop \over \akt^2} \log\left[{ 4 \prop-\akt^2
\over 2 \prop} \right]\right\}\nonumber\\
dA_{T(PP)}&=&{-2\ \log\left({4\prop-\akt^2 \over 4\prop}\right)\Phi(l^2)\over \akt^4\
Q^2}{dl^2\ d\akt^2 \over (2\pi)^3}
\label{PP}\end{eqnarray}
with $ \beta_+={ 2\sqrt{\lambda^2 + 4m_q M_V}\lambda - 2\lambda^2 - 4m_q
M_V+ M_V^2\over M_V^2}\label{betabounds}$ and
\begin{equation}
\prop=2l^2+{m_V^2-\mu_q^2\over 2}+i\epsilon
\end{equation}

We  see that the principal parts behave like $ A_L \propto {1 \over Q^3}$,
and $ A_T \propto {1 \over Q^2}$ at high $Q^2$, plus logarithmic
corrections, whereas the imaginary parts behaved like $ A_L \propto {1
\over Q^3}$, and $ A_T \propto {1 \over Q^4}$. Because this effect
is suppressed by the fall-off of the vertex function, it sets in only at
relatively large $Q^2$.

The leading behaviour of the principal parts integrals
comes from quark off-shellnesses bigger than the constituent quark mass.
Whereas in the massless case \cite{Collins} an increase in the cross section
can only come from extremely small off-shellnesses, we find that the dominant
region in the massive case is shifted by the quark mass: the principal part
integral cancels as long as one is very close to the pole, and the kinematics
is such that this cancellation is not present anymore once the off-shellness is
of the order of the quark mass squared.

Hence the source of the plateau observed at HERA is the interplay between the
principal part and the pole singularity of the amplitude.
Our model in fact
predicts that asymptotically the transverse and the longitudinal cross sections
first become equal, and that ultimately the process is dominated by the
transverse cross section. This prediction is however driven by the details of
the quark propagators, which could be modified by confinement effects.

\section{A model}
Once the cure of the problem has been identified, one can try to make a
complete model for the process. Clearly, the vertex function is not known, and
we
assume \cite{CR2} a simple behaviour, similar to that of a $1s$ wavefunction:
\begin{equation}
\Phi\left(l\right)=N\ e^{-{{\bf L}^2 \over 2 p_f^2}}
\label{L}
\end{equation}
where ${\bf L}^2$ is the quark 3-momentum in the meson rest frame, equal to
${\bf L}^2=({l.V \over M_V})^2-l.l$, and where  the Fermi momentum $p_F$ is
0.3 GeV in the $\rho$ and $\Phi$ cases, and 0.6 GeV in the J/$\psi$ case.
Furthermore, we choose the following values for the quark masses: $m_u=m_d=0.3$
GeV, $m_s=0.45$ GeV and $m_c=1.5$ GeV.
\begin{figure}[t]
\vglue 1cm
\hglue 3.7cm (a)\hglue 4.3cm (b)\\
\vglue -1.5cm
\centerline{
\hbox{\psfig{file=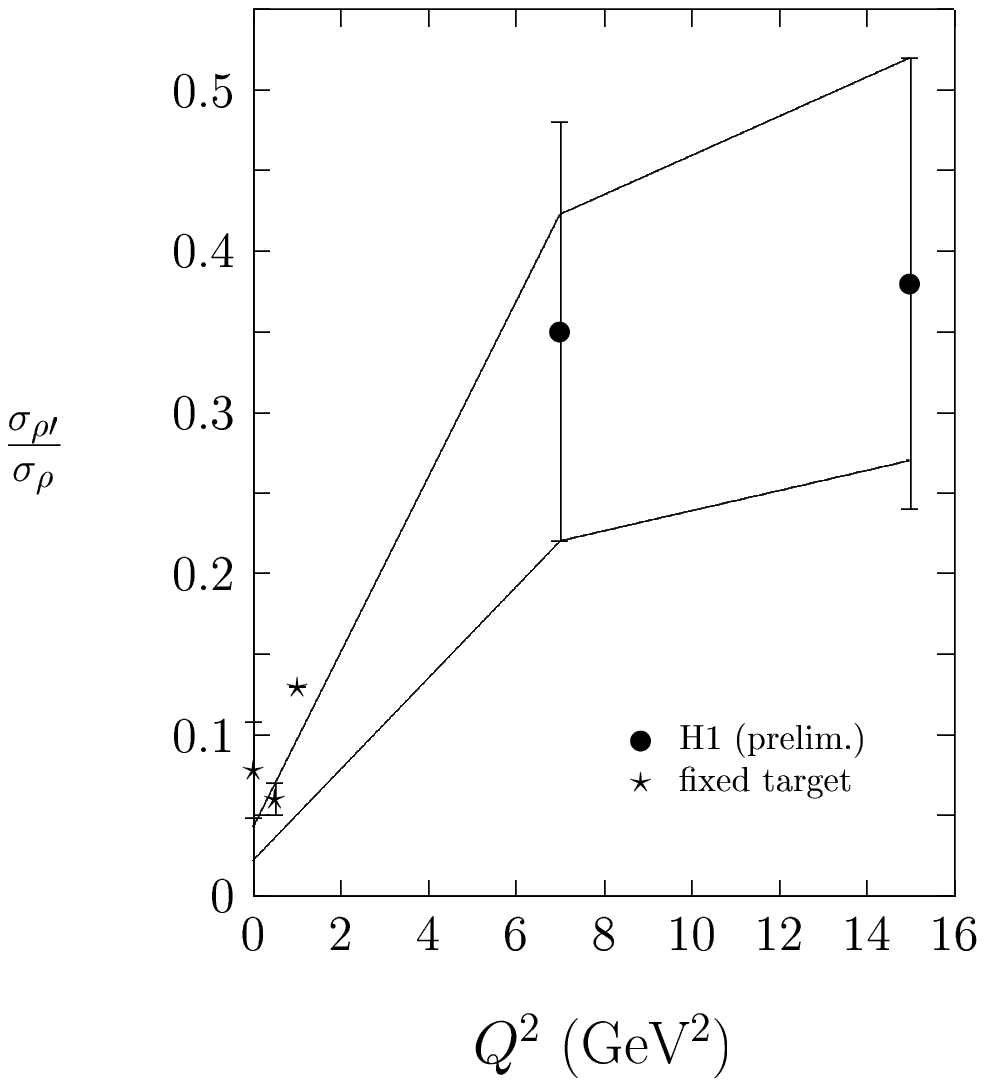,height=1.9in}
\hglue 1cm \psfig{file=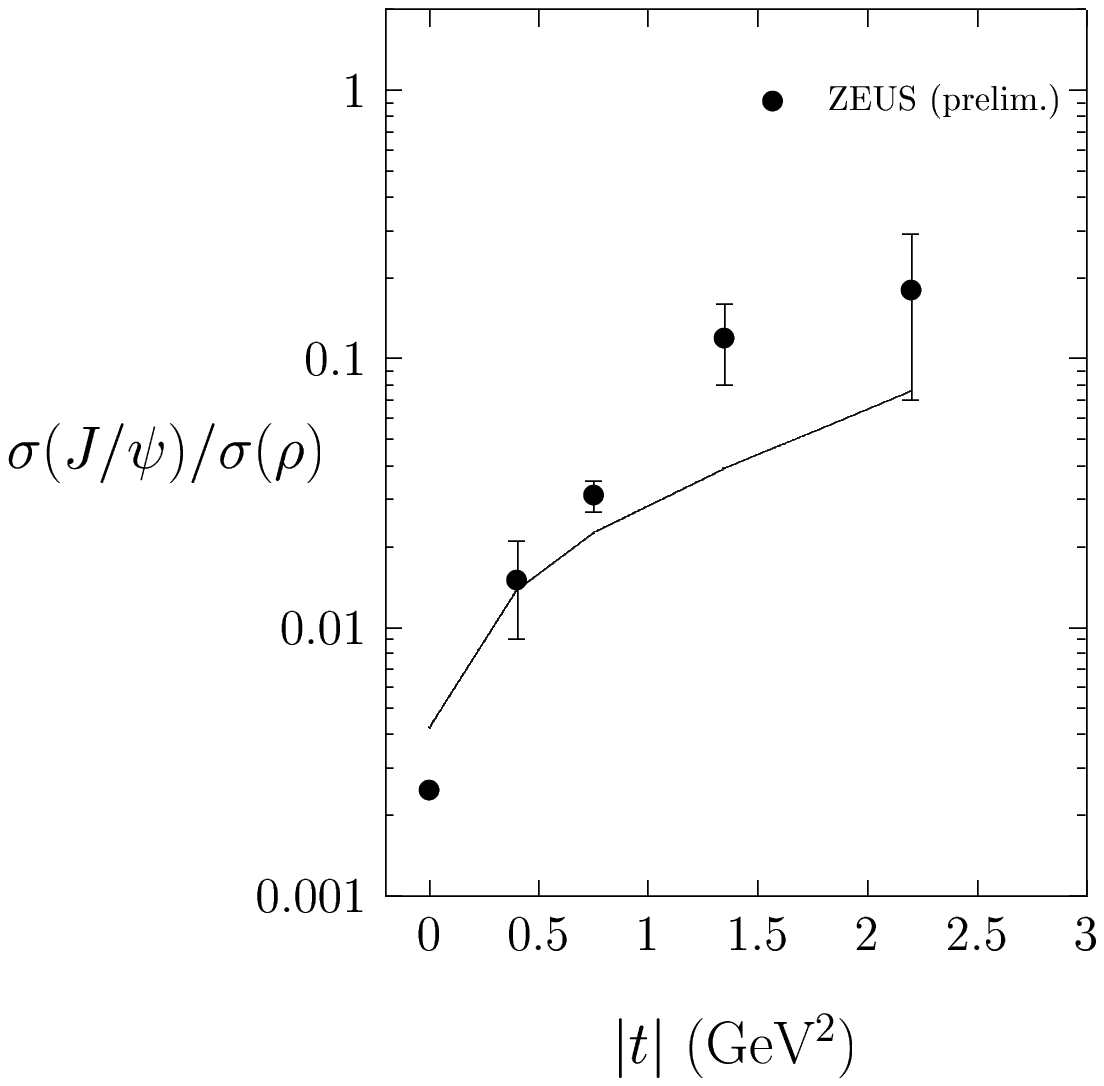,height=1.9in}}}
\caption{(a) Ratio of the $\rho'$ production cross section to that
of $\rho$ as a function of $Q^2$;\break
(b) Dependence of the ratio of the $J/\psi$ to the $\rho$ photoproduction cross
sections as a function of $t$.
}
\end{figure}
One can then normalise this vertex to reproduce the leptonic decay widths
of the various mesons. Using our previous model \cite{CR1} for the proton
form factor, we have shown that the following features are reproduced up
to an $s$-dependent factor:\\
$\bullet$ The helicity dependence of the process. This is a general feature
of all the models which use a meson vertex proportional to $\gamma_\mu$.\\
$\bullet$ The vector-meson-mass dependence of the process, in particular the
model naturally predicts the cross-over of the J/$\psi$ cross section
with that of the $\rho$ at large $Q^2$. This feature is true whether
one includes the off-shell contribution or not. \\
$\bullet$ The $Q^2$ dependence both of the longitudinal and transverse
cross sections is reproduced once the off-shell quark contribution
is included (See Fig. 1.a). In particular, the ratio $R_{L/T}$ is
easily reproduced for the typical values of $m_q$ and $p_F$ mentioned
above.\\
$\bullet$ The higher-state cross sections ($\rho'$, $\psi'$) can be reproduced
by taking a $2s$ form for the vertex. The rapid rise of the ratio of the $2s$
to the $1s$ cross sections with $Q^2$ is predicted, as can be seen in
Fig.~3.a.\\
$\bullet$ The model unexpectedly works in photoproduction. Despite the fact
that there is no hard scale present there, the shape of the $t$ distribution
comes out correctly, and the (Regge) factor multiplying the cross section
is the same as in the high-$Q^2$ case. Similarly, the ratios of the $Q^2$=0
cross sections $\sigma_\Phi/\sigma_\rho$ and $\sigma_{J/\psi}/\sigma_\rho$
as functions of $t$ turn out to be compatible with the data, as shown in
Fig.~3.b.
\\
$\bullet$ The $Q^2$ dependence of the $t$ slopes of the $\rho$ exclusive
production
cross sections is well reproduced.
\\
\section{Conclusion}
We have shown that most of the salient features of vector meson exclusive
production can be understood, provided one takes into account both the
on-shell and the off-shell contributions to the production amplitude.

One
property which is not included in this model is the energy dependence.
We find that all data are reproduced within a factor, and that this factor does
not depend on $Q^2$, on $t$ or on the meson mass. Hence on the one hand,
no BFKL enhancement at large scale is needed to agree with experiment.
On the other hand, we also find that no pomeron slope is needed either, and
hence we seem to contradict H1 measurements \cite{data}. It must be kept in
mind however
that the prediction of the $t$ dependence is the weakest point of the model so
far.  It depends
both on the proton form factor, and on the details of the infrared behaviour of
the gluon propagator \cite{JRC}, and the uncertainties affecting these may
be large enough to allow for a standard soft pomeron slope.

Nevertheless, such modifications will not affect the behaviour of the
upper quark bubble, and hence the ratio $R_{L/T}$ will remain stable.
Hence, we believe that it is now possible to
estimate reliably the contribution of the transverse cross section
and to use the total cross section measurements directly to extract
the off-diagonal gluon density.

\end{document}